\documentclass[10pt, conference, compsocconf]{IEEEtran}
\ifCLASSINFOpdf
\else
\fi
\usepackage[cmex10]{amsmath}
\usepackage{latexsym}
\usepackage{amsfonts}
\begin{document}

\newtheorem{definition}{Definition}
\newtheorem{theorem}{Theorem}[section]
\newtheorem{lemma}[theorem]{Lemma}
\newtheorem{proposition}[theorem]{Proposition}
\newtheorem{corollary}[theorem]{Corollary}

%
\title{Improving Recommendation Quality by Merging\\ Collaborative Filtering and Social Relationships}


\author{
\IEEEauthorblockN{Pasquale De Meo\IEEEauthorrefmark{1}, 
									Emilio Ferrara\IEEEauthorrefmark{4}, 
									Giacomo Fiumara\IEEEauthorrefmark{1}, 
									Alessandro Provetti\IEEEauthorrefmark{1},\IEEEauthorrefmark{7}}
\IEEEauthorblockA{\IEEEauthorrefmark{1}Dept. of Physics, Informatics Section. \IEEEauthorrefmark{4}Dept. of Mathematics. University of Messina, Italy.\\
									\IEEEauthorrefmark{7}Oxford-Man Institute, University of Oxford, UK.\\
									\{pdemeo, eferrara, gfiumara, ale\}@unime.it}
}

%


\maketitle

\begin{abstract}
Matrix Factorization techniques have been successfully applied to raise the quality of suggestions
generated by Collaborative Filtering Systems (CFSs). Traditional CFSs based on Matrix Factorization
operate on the ratings provided by users and have been recently extended to incorporate demographic
aspects such as age and gender. In this paper we propose to merge CFS based on Matrix Factorization
and information regarding social friendships in order to provide users with more accurate
suggestions and rankings on items of their interest. The proposed approach has been evaluated on a
real-life online social network; the experimental results show an improvement against existing
CFSs. A detailed comparison with related literature is also present.
\end{abstract}


%
\IEEEpeerreviewmaketitle

\section{Introduction} \label{sec:intro}

The term Collaborative Filtering (CF) refers to a wide range of algorithmic techniques targeted
at learning users' preferences to recommend them items (like commercial products or movies)
that are potentially relevant to their tastes \cite{adomavicius2005toward}.
CF techniques are an effective tool to support Web users in finding contents of their interest and,
at the same time, to limit the amount of (often useless) information they receive when
looking for recommendations.

CF techniques assume that users are allowed to rate available items over some scale and that the ratings are stored in a
{\em user-rating matrix} $\mathbf{R}$. Generally, CF techniques work by comparing the ratings provided by users.
Users who in the past have given similar ratings to the same objects can be used to give reliable estimates of a user's experience.
The rating a user $u_x$ would assign to an item $i_j$ is predicted by considering the ratings to $i_j$
provided by the users who are most similar to $u_x$ in terms of ratings.

In the latest years, many researchers have been sought to improve the accuracy of CF techniques. In
this scenario, {\em Non-Negative Matrix Factorization} techniques (in short, MF)
\cite{lee1999learning} have been largely and successfully applied to CF
\cite{koren2011advances,koren2009matrix,pilaszy2010fast}.

MF techniques take the user-rating matrix $\mathbf{R}$ as input and factorize it into the product
of two matrices $\mathbf{P}$ and $\mathbf{Q}$ in such a way as to any row of $\mathbf{P}$ (resp.,
$\mathbf{Q}$) is associated with a user (resp., an item).
Such a mapping has a nice interpretation that can be clarified through an example: assume to
consider the movie domain and assume that each user is allowed to rate movies. MF techniques map
the space of users (resp., movies) onto a {\em low dimensionality space} in which any dimension can
be conceptually interpreted as a movie genre (like ``drama'' or ``comedy'').

Once such a mapping has been performed, each user is identified by a vector whose components
specify how much she is interested in a given movie genre.

This has beneficial effects to raise the quality of produced recommendations. For instance, let us
consider two users $u_1$ and $u_2$ and assume that both of them are strongly interested {\em only}
in adventure movies and dislike other genres; finally assume that {\em none} of the movies watched
by $u_1$ have been watched by $u_2$ and vice versa (even though this setting could appear unrealistic, in real cases it is quite common).

In such a case, the comparison of the rating histories of $u_1$ and $u_2$ would not be effective:
in fact, since the overlap of the movies highly rated by $u_1$ and $u_2$ is empty, the movies liked
by $u_1$ are deemed as not relevant to $u_2$ and vice versa.

Such a conclusion is, of course, counterintuitive because both $u_1$ and $u_2$ are interested to
adventure movies and, therefore, it is likely that some of the movies liked by $u_1$ could be of
interest to $u_2$, and vice versa.
By contrast, if we would use MF techniques, we would compare two users on the basis of the genres
they like, rather than on their past rating histories and, therefore, we would not incur in the
mistakes described above.

Original MF techniques consider only user ratings and, after this, they have been extended in a
range of directions: for instance, the approach of \cite{koren2011advances} considers different rating styles
(i.e., the fact that some users tend to be more generous than others in assigning ratings) whereas
other approaches incorporate demographic information about users, like gender or age
\cite{seroussi2011personalised}.
Experimental trials show that the usage of auxiliary information is useful to produce more and more
accurate recommendations \cite{koren2011advances,seroussi2011personalised}.

In our opinion, the approaches mentioned above could be extended in such a way as to consider the
{\em social nature of the Web} and the {\em thick fabric of social relationships among Web users}.
In fact, the emergence of collaborative platforms like Facebook or Flickr encouraged users to
socialize in multiple ways: for instance, users join popular social networks like Facebook and
spend the most of their time by interacting with their friends, sharing contents like photos or
videos, and posting comments/reviews.

We believe that {\em social relationships} can provide an effective tool to raise the accuracy of
the recommendation process. This corresponds to an intuitive idea: in real life, people often ask
advices to their friends to take a decision and these advices play a crucial role in their final
decisions.

The process of requiring the help of other people to produce recommendations is not new in Computer
Science. In fact, several authors introduced the concept of {\em trust} between users and suggested
to use trust values in conjunction with CF techniques to generate suggestions
\cite{massa2004trust,golbeck2006inferring,o2009capturing}.

In this paper we propose to consider social relationships between users in conjunction with user
ratings to generate recommendations. Our approach relies on MF techniques, but introduces some
novel contributions.

First of all, differently from traditional approaches, our approach merges both information about
user ratings and information on social relationships. In detail, our approach to learning the
latent space relies on the idea that if two users are tied by a social relations like
friendship, then they should be mapped onto ``close'' vectors in the latent space. This has a
relevant consequence: opinions/ratings of friends of a user $u_x$ will be more influential
than opinions/ratings provided by users who are unknown to $u_x$.

As a further contribution, our approach suggests to use social relationships instead of trust ones.
This provides three main novelties: the first is that trust relationships are generally {\em
asymmetric}, in the sense that if a user trusts another one, the opposite may not hold true. By
contrast, social relationships can be both {\em symmetric} (e.g., friendship relations) and
{\em asymmetric} (e.g., a user $u_x$ who posts a comment on the blog of a user $u_y$ but $u_y$
never replied $u_x$).
As a second novelty, users are required to explicitly declare what users they trust: for instance,
in a system like Slashdot\footnote{{\tt http://slashdot.org}} users are allowed to declare their
friends and foes. As for social relationships, we can use {\em explicit} information provided by
the users but we can also {\em unobtrusively} monitor user behaviors to learn her preferences and
her relationships with others. For instance, we can analyze the comments posted by multiple users
on a given subject (like the review of a commercial product) to learn the personal opinion of a
particular user as well as to identify pairs of users showing divergent (resp., convergent)
opinions.

Finally, as for the third novelty, we observe that the definition of trust relationships relies on
the fact that a user explicitly assumes that her interactions with another one are beneficial to
her. By contrast, in social relationships, there are many reasons driving two users to interact and
some of them {\em do not necessarily imply} that a user get some benefit from another one: for
instance, a user could get in touch with another only to expand her knowledge on a topic.

The plan of the paper is as follows: in Section \ref{sub:matrixfact} we review MF techniques for
Collaborative Filtering. In Section \ref{sec:ourapproach} we describe our approach. In Section
\ref{sec:experiments} we illustrate the experiments we carried out to validate our technique.
Related literature is discussed in Section \ref{sec:related}. Finally, in Section
\ref{sec:conclusions} we draw our conclusions.

\section{Matrix Factorization for Collaborative Filtering} \label{sub:matrixfact}

In the latest years, many researchers proposed to apply {\em dimensionality reduction} (DR)
techniques to increase the accuracy of Collaborative Filtering (CF) techniques.

In detail, a CF algorithm operates on a matrix $\mathbf{R}$ called {\em user-rating} matrix. The
$\mathbf{R}$ matrix has $n_u \times n_i$ entries, being $n_u$ (resp., $n_i$) the number of users
(resp., items); its generic entry $\mathbf{R}_{xj}$ is the rating the user $u_x$ applied on the
item $i_j$. We set $\mathbf{R}_{xj} = ``\star''$ if the user $u_x$ has not rated $i_j$. The matrix
$\mathbf{R}$ is {\em quite sparse} because in real-life domains, the number of available items is
usually very large in comparison with the number of items typically evaluated by a user. Data
sparsity is one of the major drawbacks plaguing CF algorithms; in fact, it negatively affects the
computation of similarities between users and items and this, ultimately, yields poor results in
predicting unknown ratings \cite{sarwar2000analysis}.

DR techniques have proven to be effective in fighting against data sparsity
\cite{sarwar2000analysis}. The key idea of DR techniques is to {\em map the user-rating} matrix
onto a {\em latent space} of dimension $k$, being $k$ a fixed integer. Due to this mapping, an item
$i_j$ is represented by an {\em item vector} $\mathbf{q}_j \in \mathbb{R}^k$ and a user $u_x$ is
associated with a {\em user vector} $\mathbf{p}_{x} \in \mathbb{R}^k$.

A dimension in the latent space identifies a feature describing an item or a user; for instance, if
we consider the movie domain, the dimensions may be interpreted as the genres of a movie.
For a given item $i_j$, the generic entry $\mathbf{q}_j[\ell]$, $0 \leq \ell \leq k$ of the item
vector specifies whether $i$ possesses the $\ell$-th factor and the strength of this possession. An
analogous interpretation holds for user vectors.

To better clarify these concepts, let us consider again the movie domain; a possible latent space
associated with it would consist of movie genres like ``comedy,'' ``adventure,'' ``romance'' and
``horror.''
An item $i_j$ could be represented as $\mathbf{q}_i = [ 0.5, 0.7, 0, 0]$ specifying that the genre
of $i_j$ is a blend of ``comedy'' and ``adventure.''

In the latent space, the approval rating of a user $u_x$ to an item $i_j$ is computed as
the inner product $\mathbf{p}_x \cdot \mathbf{q}_j$.

A possible approach to learning the latent space is given by {\em Singular Value Decomposition
(SVD)}. The SVD approximates $\mathbf{R}$ by means of its eigenvalues and it has been widely and
successfully applied in the Information Retrieval context
\cite{manning2008introduction,martin2011extraordinary}. Unfortunately, in the context of CF, the
matrix $\mathbf{R}$ is {\em not only sparse} but many of its entries are also {\em unspecified},
i.e., there are a lot of entries labeled with the symbol $``\star''$ because many users may not be
aware about the existence of an item and, therefore, we can not conclude that she likes/dislikes
the item itself. In such a configuration, extensive experimental trials show that SVD is able to
achieve poor performance \cite{koren2009matrix}.

Recently, many researchers suggested to apply {\em matrix factorization techniques} to CF systems
\cite{koren2009matrix,koren2011advances,pilaszy2010fast,seroussi2011personalised}. In this case, the computation of the latent space
requires to solve a suitable {\em optimization problem}.

More formally, let $\mathbf{P}$ (resp., $\mathbf{Q}$) be a $n_u \times k$ (resp., $n_i \times k$)
matrix such that the $x$-th row $\mathbf{p}_x$ of $\mathbf{P}$ represents the vector associated
with the user $u_x$; in an analogous fashion, the $j$-th row $\mathbf{q}_j$ of $\mathbf{Q}$
represents the vector associated with the item $i_j$ .
The optimization problem to solve is
\begin{equation}
\label{eqn:optprob}\min_{\langle \mathbf{P},\mathbf{Q} \rangle} {\cal L} = \min \frac{1}{2} ||\mathbf{R} - \mathbf{P}\mathbf{Q}^T||^2_{F} +
\frac{\lambda}{2} \left(||\mathbf{P}||^{2}_F + ||\mathbf{Q}||^{2}_F\right)
\end{equation}

Here $\mathbf{Q}^T$ is the transpose of the matrix $\mathbf{Q}$ whereas the symbol $||\cdot||_{F}$
denotes the {\em Frobenius} norm of a matrix\footnote{The {\em Frobenius norm} of a matrix
$\mathbf{A}$ is defined as $||\mathbf{A}||_F = TR\left(\mathbf{A}\mathbf{A}^T\right)$ being $TR$
the trace of the matrix $\mathbf{A}$, i.e., the sum of the elements located on its main diagonal.}.
Finally, the term $\frac{\lambda}{2} \left(||\mathbf{P}||^{2}_F + ||\mathbf{Q}||^{2}_F\right)$ is
known as {\em Tikhonov regularization} and it is used to avoid {\em overfitting}. The parameter
$\lambda$ is usually computed by applying {\em cross-validation}.

A popular strategy to optimize the function ${\cal L}$ is based on the so called {\em gradient
descent method} \cite{lee1999learning}. To implement such a strategy, we must consider {\em only} the
ratings actually provided by the users and ignore missing entries in $\mathbf{R}$. To this purpose,
if we set $\delta_{xj} = 1$ if $u_x$ has rated $i_j$ and 0 otherwise, the optimization problem to
solve is
\begin{equation}
\label{eqn:eqn_known_ratings}
\min \frac{1}{2} \sum_{x = 1}^{n_u}\sum_{j = 1}^{n_i}\delta_{xj} \left(\mathbf{R}_{xj} - \mathbf{p}_x \cdot \mathbf{q}_j \right)^2 + \frac{\lambda}{2}
\left(||\mathbf{P}||^{2}_F + ||\mathbf{Q}||^{2}_F\right)
\end{equation}

The gradient descent procedure consists of four steps:

\begin{enumerate}

\item The vectors $\mathbf{p}_x$ and $\mathbf{q}_j$ are initialized at random.

\item The partial derivatives of ${\cal L}$ are computed.

\item The vectors $\mathbf{p}_x$ and $\mathbf{q}_j$ are updated in the direction opposite to
    the partial derivatives
    $$
    \mathbf{p}_x^{'} = \mathbf{p}_x - \beta \frac{\partial {\cal L}}{\mathbf{p}_x} \qquad
    \mathbf{q}_j^{'} = \mathbf{q}_j - \beta \frac{\partial {\cal L}}{\mathbf{q}_j}
    $$

Here $\beta$ is a threshold to be determined.

\item Steps 2-3 are iterated until a particular number of iterations has been carried out or
    the improvement of the function ${\cal L}$ is less than a given threshold $\varepsilon$.

\end{enumerate}

\section{Approach Description} \label{sec:ourapproach}

In this section we describe our approach to merge social relationships with user ratings. We assume
that users can create various type of social relationships (e.g., getting friends or affiliating to
the same groups). We say that a {\em social tie} exists between two users if they created a social
relationship. A social tie can be {\em symmetric} or {\em asymmetric}. To introduce our approach,
we need the following definitions:

\begin{definition}
Let $U = \{ u_1, u_2, \ldots, u_{n_u}\}$ be a set of users and $I = \{ i_1, i_2, \ldots,
i_{n_i}\}$ be a set of items. A {\em Social Rating Network (SRN}) is a 4-tuple $SRN = \langle U, I,
\phi, \psi \rangle$ where:

\begin{itemize}

\item $\phi: U \times I \rightarrow \mathbb{R}^{+} \cup \{``\star''\}$ is a function (called
    {\em rating function}) which associates a user $u_x \in U$ and an item $i_l \in I$ with a
    real number $r_{xl}$ if $u_x$ rated $i_l$ with $r_{xl}$, and with the symbol $``\star''$ if
    $u_x$ has not rated $i_l$.

\item $\psi: U \times U \rightarrow\{ 0, 1 \}$ is a function (called {\em social function})
    which takes a pair of users $u_x, u_y \in U$ as input and returns 1 if and only
    if a {\em social tie} exists between them and 0 otherwise.

\end{itemize}

\end{definition}

Due to the definition of social tie, the function $\psi$ is generally {\em asymmetric}, i.e.,
$\psi(u_x,u_y) \neq \psi(u_y,u_x)$, for some pairs of users $u_x$ and $u_y$. The function $\psi$ is
also useful to define the concept of {\em neighborhood} of a user $u_x$ in an SRN:

\begin{definition}
Let $SRN = \langle U, I, \phi, \psi \rangle$ be a Social Rating Network and $u_x \in U$ be a
user. The {\em neighborhood} $N_x$ of $u_x$ is defined as the set of users having a social tie with
$u_x$
$$
N_x = \{u_y \in \mathcal{U} : \psi(u_x,u_y) = 1\}
$$

\end{definition}

The concept of SRN specializes the concept of social network; in fact, unlike traditional social
networks, users are not only allowed to {\em interact} among each others but also to {\em rate}
objects.
To describe our approach, we start observing that Equation \ref{eqn:optprob} provides an effective
tool for tackling data sparsity but it is {\em agnostic} about social relationships because no
terms related to social ties involving existing users appear in it.

However, in reality, social relationships are a powerful tool for producing suggestions: for
instance, if a person is uncertain about the purchase of a good, she often asks her
friends/acquaintances opinions or advices that often play a crucial role in her final
decision.

Our goal is, therefore, to {\em extend} Equation \ref{eqn:optprob} to the case of SRNs by adding
terms capable of taking into account social ties among users. To simplify the discussion, henceforth we shall
focus on friendship relations as models of social relationships; however, without any
loss of generality, the conclusions we will draw are still valid for other type of social
relationships.

We start observing that, if we would solve the optimization problem in Equation \ref{eqn:optprob},
{\em we would map any user} $u_x$ onto a {\em point} $\mathbf{u}_x$ in the latent space.
We guess that, if two users are friends, then they should be mapped onto {\em close points}, in the
latent space; in other words, given three users $u_x$, $u_y$ and $u_z$, such that only $u_x$ and
$u_y$ are friends, the distance between the points $\mathbf{p}_x$ and $\mathbf{p}_y$ must be less
than the distance between $\mathbf{p}_x$ and $\mathbf{p}_z$.

Such a requirement has an easy explanation: in fact, if a user $u_x$ is close to her friends in the
latent space, the opinions of the friends of $u_x$ will be more relevant to recommend items to
$u_x$ than the opinions of other users who are unknown to $u_x$.
From a mathematical standpoint, the distance between two users $u_x$ and $u_y$ is given by
$||\mathbf{p}_x - \mathbf{p}_y||$, being $||\cdot||$ the euclidian norm in the latent space. If we
denote as $N_x$ the neighborhood of $u_x$, our aim is that the term $\sum_{u_y \in N_x}
||\mathbf{p}_x - \mathbf{p}_y||$ is as small as possible.

These considerations suggest to add a {\em penalty term} to Equation \ref{eqn:optprob}. In detail,
among the all possible mappings to the latent space, we decide to penalize those mappings in which
users tied by a friendship relation are mapped onto ``far'' points.
In the light of these considerations, the optimization problem to solve is as follows

\begin{equation}
	\begin{split}
		\min_{\langle \mathbf{P},\mathbf{Q} \rangle} {\cal L}^{'} = \min \frac{1}{2} ||\mathbf{R} - \mathbf{P}\mathbf{Q}^T||^2_{F} + & \\
		+ \lambda \left(||\mathbf{P}||^{2}_F + ||\mathbf{Q}||^{2}_F\right) + \mu \sum_{x = 1}^{n_u} \sum_{u_y \in N_x} || \mathbf{p}_x - \mathbf{p}_y||
	\end{split}
\label{eqn:optprobsoc}
\end{equation}

As in the previous case, $\lambda$ and $\mu$ are two constants to be tuned to avoid overfitting.
We applied the gradient descent algorithm to solve our optimization problem. In such a case, the
partial derivatives of the objective function ${\cal L}^{'}$ are
$$
\frac{\partial {\cal L}^{'}}{\partial \mathbf{p}_x} = \sum_{j = 1}^{n_i} \delta_{ij} \left(\mathbf{p}_x \cdot \mathbf{q}_j - \mathbf{R}_{xj} \right)
\mathbf{q}_j + \lambda \mathbf{p}_x + \mu \sum_{u_y \in N_x} \left( \mathbf{p}_x - \mathbf{p}_y  \right)
$$

$$
\frac{\partial {\cal L}^{'}}{\partial \mathbf{q}_j} = \sum_{x = 1}^{n_u} \delta_{ij} \left(\mathbf{p}_x \cdot \mathbf{q}_j - \mathbf{R}_{xj} \right)
\mathbf{p}_x + \lambda \mathbf{q}_j
$$

\section{Experiments} \label{sec:experiments}

In this section we present the experiments we carried out to evaluate our approach.
We built a social networks (called {Cofe}) in which users were allowed to rate movies. The early
users of Cofe were students enrolled in a BsC degree in Computer Science at our University; after
this, students were allowed to invite their friends to join Cofe, create friendship relationships
with other members, insert movie titles and rate them.
We gathered data on 37 students and 297 movie ratings. We used as metric to assess the quality of
recommendations the {\em Root Mean Square Error} (RMSE); to define it, assume to {\em randomly}
split the dataset in two parts called {\em training} and {\em test set}. The training set is used
to perform matrix factorization whereas the test set is used to assess the quality of
recommendations. If we define {\em (i)} $r_{xj}$ the rating $u_x$ assigned to $i_j$, {\em (ii)}
$\hat{r}_{xj}$ the rating predicted by a given method for $u_x$ and $i_j$ and {\em (iii)} $N_t$ the
size of training set, the RMSE is defined as $ RMSE = \sqrt{\frac{1}{N_t} \left(r_{xj} -
\hat{r}_{xj}\right)^2}$.

We compare our approach with the following methods:

\begin{itemize}

\item {\em Naive}. In this method, the rating of an item is predicted as the average of the
    ratings provided by all users.

\item {\em NMF}. In this method, we solve the optimization problem in Equation
    \ref{eqn:optprob} to predict missing ratings.
\end{itemize}

To run both {\em NMF} and our approach (hereafter, {\em OA}) we fixed, after a pre-tuning activity,
$\lambda = 0.001$. In addition, we fixed the number of iterations of both {\em NMF} and {\em OA}
equal to 5000.

In a first experiment we studied how the value of $k$ (i.e., the number of dimensions in the latent
space) impacted on the accuracy. We fixed $k= \{2, \ldots, 16\}$ and ran both {\em NMF} and {\em
OA}; of course, the {\em Naive} algorithm is not influenced by $k$.
The obtained results are reported in Table \ref{tab:RMSEk}.

\begin{table}[t]
    \centering
    \footnotesize
    \begin{tabular}{||c|c|c|c||}
    \hline \hline
    {\em k}          & \multicolumn{3}{|c|}{RMSE}\\
    \hline \hline
                     & {\em Naive}       & {\em NMF}        & {\em OA}\\
    \hline
    2        &  0.897     &   0.825     & 0.816  \\
    \hline
    4        &  0.897      &   0.751         & 0.742   \\

    \hline
    6        &  0.897     &   0.739 &   0.731\\
    \hline
    8        &  0.897      &  0.727&   0.719\\

    \hline
    10        &  0.897     &  0.71&  0.705 \\
    \hline
    12        &  0.897      & 0.715&  0.707\\

    \hline
    14        &  0.897     &  0.717 &  0.713\\
    \hline
    16        &  0.897      &  0.717&  0.714\\

    \hline \hline
    \end{tabular}
    \caption{RMSE of {\em Naive}, {\em NMF} and {\em OA} results for different values of $k$}
    \label{tab:RMSEk}
    \vspace{-.5cm}
\end{table}

From the analysis of this table we can conclude that:

{\em - OA} significantly outperforms the {\em Naive} method and is better than {\em NMF}.

{\em - OA} and {\em NMF} achieve their best performance when $k$ is around 10. In fact, if
    the number $k$ of dimensions exploited to represent user preferences and movie features is
    too low, it is not possible to accurately capture the differences between two movies or two
    users and, therefore, it is hard to correctly recommend movies to users.

    By contrast, if $k$ exceeds 10, the RMSE of both {\em OA} and {\em NMF} does not
    significantly decrease. This means that a number of latent factors equal to 10 is enough to
    correctly capture movie genres and a further increase of $k$ is useless.

    However, the higher $k$, the larger the size of $\mathbf{P}$ and $\mathbf{Q}$ and,
    therefore, the larger the space of memory required to store them.
    These considerations indicate us that a good trade-off between space requirements and
    recommendation accuracy is achieved when $ k = 10$.

As a further experiment, we tuned the value of $\mu$, i.e., we studied how the weight associated
with social relationships reflect on the quality of suggestions.
The RMSE value (when $k = 10$) and $\mu$, ranged from $10^{-6}$ to $10^{-1}$, are reported in Table \ref{tab:mu}.
\begin{table}[t]
    \centering
    \footnotesize
    \begin{tabular}{||c|c||}
    \hline \hline
    $\mu$          & {\em RMSE}\\
    \hline \hline
    $10^{-1}$        &  0.695\\
    \hline
    $10^{-2}$        &  0.68\\

    \hline
    $10^{-3}$        &  0.675\\
    \hline
    $10^{-4}$        &  0.687\\

    \hline
    $10^{-5}$        &  0.707\\
    \hline
    $10^{-6}$        &  0.71\\

    \hline \hline
    \end{tabular}
    \caption{Impact of $\mu$ on RMSE}
    \label{tab:mu}
    \vspace{-.5cm}
\end{table}
From Table \ref{tab:mu} we observe that if $\mu \rightarrow 0$, the term representing social
relationship tends to vanish and, therefore, {\em OA} degenerates into {\em NMF} (i.e., only user
ratings are considered). By contrast, if $\mu \rightarrow 10^{-1}$, an opposite effect arises:
information on social relationships dominates over user ratings. The best value of RMSE is achieved
if $\mu $ is around $10^{-3}$ because, in such a configuration, our approach takes advantage of
both user ratings and social relationships.
%

\section{Related Works} \label{sec:related}

In this section we describe some approaches related to our research. In detail, we first
consider Matrix Factorization (MF) techniques and describe how they have been applied in the
context of Collaborative Filtering (CF); after this, we highlight the main novelties introduced by
our approach. Then, we focus on {\em Trust-Based Collaborative Filtering Systems}, i.e., on CF
systems in which users are allowed to explicitly declare if they trust (and sometimes distrust)
other users. We explain the differences between trust-based CF systems and our research
efforts.

\subsection{Matrix Factorization Techniques for CF}
\label{sub:matrixfactorization}

The notion of Non-Negative Matrix Factorization (in short, MF) was introduced for the first time in
the seminal paper of Lee and Seung in 1999 \cite{lee1999learning}. MF techniques have been widely applied in
multiple domains like data clustering \cite{ding2005equivalence} and bioinformatics \cite{jung2008application}.

The algorithm proposed in \cite{lee1999learning} to perform matrix factorization is {\em iterative} and it
strongly resembles the gradient descent method discussed in this paper.

One of the first approaches that exploited MF in the context of Collaborative Filtering is
\cite{bell2007scalable}. In that paper, the authors observe that the task of factorizing two matrices is
computationally challenging and, therefore, they propose a strategy called {\em Alternating Least
Square (ALS)}.

ALS-based techniques proceed iteratively and each iteration consists of two stages: in the first stage
the matrix $\mathbf{P}$ is fixed and the problem described in Equation \ref{eqn:optprob} is solved
with respect to the matrix $\mathbf{Q}$. In the second stage, the vice versa.
In both the two stages, the optimization problem can be reformulated (and solved) as a {\em least
square} problem \cite{bell2007scalable,koren2009matrix}.
ALS techniques lend themselves to a massive amount of parallelization. Therefore, the growing
success of distributed computing platform like HADOOP, envisages a big success of these techniques.

As further extensions, the approach of \cite{koren2011advances} suggests to add {\em bias terms} in Equation
\ref{eqn:optprob}; these terms model the fact that some users tend to provide more generous ratings
than others or that some items tend to receive ratings generally higher than other items. In the
same paper, the authors study how temporal changes in user preferences and/or in item properties
(e.g., the fact that the popularity of an item may decay over time) impact on the process of
generating recommendations.

Some approaches target at learning the latent space in presence of {\em implicit user feedbacks}
\cite{pilaszy2010fast}. In this scenario, users do not directly provide ratings on items but the analysis
of user behaviours is useful to understand user preferences. For instance, in an e-commerce Web
site, we can assume that a user likes an item if she bought it.
Finally, some authors \cite{seroussi2011personalised} suggest to incorporate in  Equation \ref{eqn:optprob} also
{\em demographic variables} like the gender or the age of the users.


Differently from the works above, in our approach, the task of learning the latent space relies on
information tied to both {\em user ratings} and {\em their social relationships}. In other words,
we merge information about social relationships and user behaviours (encoded as the ratings they
provided) in a {\em unique framework} to produce recommendations.

\subsection{Trust-based CF systems}
\label{socialnetworkdata}

The usage of trust information to generate recommendations has been largely explored in the
literature.

Some approaches suggest to use trust values between pairs of users in conjunction with
Collaborative Filtering techniques to produce recommendations
\cite{MeoNQRU09,golbeck2006using,massa2004trust}.

The key challenge in trust-based approaches is represented by the computation of trust values. In
fact, in most cases, users provide few explicit declarations of trust; this implies that trust
values among many pairs of users are unknown and, therefore, a mechanism for inferring new trust
values from existing ones must be designed.

Some approaches assume that existing users form a community which is modeled through a graph $G$
whose nodes represent users and edges indicate relationships between them \cite{golbeck2006inferring,guha2004propagation}.
In \cite{golbeck2006inferring} the authors apply a modified version of the Breadth First Search algorithm on $G$
to infer multiple values of trust for each user; the average of these values is computed to produce
a final trust value.

The approach of \cite{guha2004propagation} considers paths up to a fixed length in $G$ and propagates trust
values explicitly declared by users on them to infer new ones.

Recently, some approaches studied the problem of propagating trust in {\em signed social networks},
i.e., social networks in which ties between users may be either {\em positive} (indicating, for
instance, that two users are friends) or {\em negative} (indicating a relationship like antagonism)
\cite{kunegis2009slashdot,leskovec2010predicting}.
These approaches are grounded in {\em balance theory} introduced in social psychology. Roughly
speaking, balance theory is based on principles like {\em the enemy of my friend is my enemy} and
{\em the friend of my enemy is my enemy}.

Differently from the approaches described above, our approach is based on {\em social
relationships}. This has two major effects: {\em (i)} Social relationships are {\em weaker} than
trust ones: in fact, in a social network two users may interact for different reasons (e.g., to be
informed on a new topic). By contrast, if a user $u_x$ trusts a user $u_y$, this means that $u_x$
has a reasonable expectation that interactions with $u_y$ will be beneficial for her. {\em (ii)}
Trust Relationships are {\em asymmetric}. On the contrary, social relationship can be asymmetric as
well as symmetric.

During the latest years, the possibility of exploiting the social relationships together with the rating behaviors has been advanced  \cite{bonhard2006knowing,said2010social}.
In \cite{bonhard2006knowing} the authors suggest that drawing on similarity and familiarity between the users in their rating activities could support the decision making and increase recommendation quality.
In \cite{said2010social} authors propose a model that combines social ties and ratings to improve the movie recommendation quality in the context of a real-world social rating network, providing encouraging results.
With this paper we additionally substantiate the hypothesis that combining Collaborative Filtering and social relationships is helpful in order to build better Recommender Systems.

\section{Conclusions} \label{sec:conclusions}
In this work we presented a novel strategy to provide heightened quality recommendations to users,
in the context of Social Rating Networks, those Social Networks in which users are allowed to
socially interact and to rate items. Our approach relies both on user ratings and on social
relationships among users and merges this information by means of Matrix Factorization techniques.
Experimental results show the effectiveness of our approach.

We plan to extend our approach to Trusted Social Networks, i.e., social networks in which users are allowed to
define positive or negative trust in other users. Another issue to explore is related to the
scalability of our approach. In detail, we plan to design and implement efficient (and possibly
distributed) algorithms to perform Matrix Factorization.





%


\bibliographystyle{IEEEtran}

\bibliography{IEEEabrv,isda2011-rec-sys-bib}

\end{document}